\documentclass{article}
\usepackage{spconf,amsmath,graphicx}

\usepackage{xcolor}
\usepackage{hyperref}
\usepackage[utf8]{inputenc}
\usepackage[T1]{fontenc}
\usepackage{footnote}
\makesavenoteenv{tabular}
\makesavenoteenv{table}

\title{BUT System for the Second DIHARD Speech Diarization Challenge}

\name{\begin{tabular}{c}Federico Landini$^1$,  Shuai Wang$^{1,2}$, Mireia Diez$^1$, Luk\'{a}\v{s} Burget$^1$, Pavel Mat\v{e}jka$^1$, Kate\v{r}ina \v{Z}mol\'{i}kov\'{a}$^1$, \\Ladislav Mo\v{s}ner$^1$, Anna Silnova$^1$, Old\v{r}ich Plchot$^1$, Ond\v{r}ej Novotn\'{y}$^1$, Hossein Zeinali$^1$, Johan Rohdin$^1$\thanks{The work was supported by Czech Ministry of Education, Youth and Sports from the National Programme of Sustainability (NPU II) project "IT4Innovations excellence in science - LQ1602".}\end{tabular}}

\address{$^1$Brno University of Technology, Faculty of Information Technology, IT4I Centre of Excellence, Czechia \\
        $^2$ Speechlab, Department of Computer Science and Engineering, Shanghai Jiao Tong University, China \\
        \textit{\{landini,mireia\}@fit.vutbr.cz}}

\begin{document}

\maketitle

\begin{abstract}
This paper describes the winning systems developed by the BUT team for the four tracks of the Second DIHARD Speech Diarization Challenge. For tracks 1 and 2 the systems were mainly based on performing agglomerative hierarchical clustering (AHC) of x-vectors, followed by another x-vector clustering based on Bayes hidden Markov model and variational Bayes inference. We provide a comparison of the improvement given by each step and share the implementation of the core of the system. For tracks 3 and 4 with recordings from the Fifth CHiME Challenge, we explored different approaches for doing multi-channel diarization and our best performance was obtained when applying AHC on the fusion of per channel probabilistic linear discriminant analysis scores.
\end{abstract}


\vspace*{-1.5mm}
\section{INTRODUCTION}
\label{sec:introduction}
\vspace*{-2mm}
With the aim of bringing attention to diarization performed on challenging data, the Second DIHARD Diarization Challenge~\cite{ryant2019second} proposed a common ground for comparison of different diarization systems on multiple domains. Alike the first DIHARD Diarization Challenge~\cite{ryant2018first}, tracks 1 and 2 consisted in performing diarization on single-channel recordings from different domains with and without oracle voice activity detection (VAD) labels, respectively. Tracks 3 and 4 focused on multi-channel data from the Fifth CHiME Challenge~\cite{barker2018fifth} also with and without VAD labels, respectively. 

Our effort allowed us to obtain the first position on all four tracks. This paper describes those four winning systems but due to the lack of space some details are omitted. For specific parameter configurations we refer the reader to \cite{landini2019but}. Together with this publication, the code of the most relevant modules of the winning system of track 1 has been made available \cite{DIHARDrecipeBUT}.

The challenge proposed four tracks working with two different datasets and this paper will be structured accordingly.
Section \ref{sec:dihard} describes the complete processing pipeline that we used for speaker diarization in tracks 1 and 2: We describe the used signal pre-processing, x-vector extraction and agglomerative hierarchical clustering (AHC) of x-vectors applied to obtain initial labels for the following step. In the next step, which is the core of our diarization pipeline, x-vectors are clustered using Bayesian hidden Markov model (BHMM). This method is often referred to as {\em VB diarization}~\cite{DiezInter19}. Next, frame-level re-segmentation is performed based on another BHMM and finally, the overlapped speech post-processing is applied.
Since we did not have oracle VAD labels for track 2, we describe how we trained a VAD system. Experiments related to each part of the system are described in each subsection followed by their corresponding discussions.

Section \ref{sec:chime} focuses on processing the CHiME data used in tracks 3 and 4. We describe the clustering method used for diarization and the experiments we carried out to take advantage of the multi-channel data. We also show the performance when using a VAD system instead of the oracle labels. 

Finally, we comment on the conclusions we reached to during and after the challenge. We also comment on the challenges that we see in the task of speaker diarization and the paths we believe we should follow.

\vspace*{-1.5mm}
\section{SYSTEMS for tracks 1 and 2, DIHARD DATA}
\label{sec:dihard}
\vspace*{-2mm}
In previous works we have shown that the BHMM-based diarization system can be tuned for different domains in order to achieve much better performance~\cite{DiezIEEE19}. Still, this requires a module capable of classifying recordings with great accuracy. Since misclassification in this step can be quite harmful, we opted for focusing on a single diarization system for all domains.
The different parts of the system 
are described and their results discussed in the following subsections. 
In this section, results in gray indicate ``cheating'' results where the test data is also used for training (i.e. the development set).

\vspace*{-1.5mm}
\subsection{Signal Pre-processing}
\label{sec:preprocessing_dihard}
\vspace*{-1mm}
Due to the high levels of noise and reverberation in some of the recordings, we explored four different enhancing methods for improving the quality of the signals. 
We compared the method provided by the organizers\footnote{https://github.com/staplesinLA/denoising\_DIHARD18}, one based on the Wave-U-Net~\cite{macartney2018improved}, one based on neural network (NN) autoencoders~\cite{plchot-enhancement-2016} and the weighted prediction error (WPE)~\cite{nakatani-wpe-2010,Drude2018NaraWPE} method which removes late reverberation. Alike the first edition of the challenge~\cite{diez2018but}, we found WPE to be the most effective in this regard. Therefore, we pre-processed all training, development and evaluation recordings using this technique.

\vspace*{-1.5mm}
\subsection{x-vectors}
\label{sec:xvecs}
\vspace*{-1mm}
A very successful recent approach for speaker diarization is to cluster deep neural network (DNN) based speaker embeddings known as x-vectors~\cite{SellSDIHARD18,DiezInter19}. From the input recordings, x-vectors are typically extracted every 0.75s from 1.5s overlapping sub-segments. For this challenge, we use a higher x-vector frame-rate of 0.25s for tracks 1 and 2 as we found it to significantly improve the results~\cite{DiezICASSP20}.
In our diarization pipeline, x-vectors are clustered in two steps using AHC and BHMM as described in the following sections.

The x-vector extractor used is based on the SRE16 recipe~\cite{snyder_kaldi_recipe} from the Kaldi toolkit~\cite{povey2011kaldi} with some modifications: We use a larger and deeper neural network for x-vector extraction, which is trained on VoxCeleb 1 and 2~\cite{chung2018voxceleb2} with data augmentation. We use more training epochs than the original recipe and a different strategy to generate the input examples from the training speech recording. More details can be found in \cite{landini2019but}.

\vspace*{-1.5mm}
\subsection{AHC Initial Clustering}
\label{sec:initialization}
\vspace*{-1mm}
The x-vectors extracted from an input recording are first clustered by means of AHC with similarity metric based on probabilistic linear discriminant analysis (PLDA)~\cite{kenny10PLDA_HTP} log-likelihood ratio scores, as used for speaker verification. 
We train two PLDA models for this purpose: The first one is trained on x-vectors extracted from 3 seconds speech segments from VoxCeleb 1 and 2 which are mean centered, whitened to have identity covariance matrix and length-normalized~\cite{GarciaRomero2011lnorm}. 
The centering and whitening transformation are estimated on the joint set of DIHARD development and evaluation data to have a better estimate since unsupervised use of the evaluation data is allowed.
To take advantage of the in-domain data, the second PLDA model is trained on x-vectors extracted in a similar way using the DIHARD development data. The centering and whitening transformation are also estimated on the joint set of DIHARD development and evaluation data. Note that these transformations are applied both to development and evaluation x-vectors when performing diarization.

The final ``domain-adapted'' PLDA model used for AHC-based clustering is obtained as an interpolation of the two PLDA models: means, within- and across-class covariance matrices from the two models are averaged.
Table~\ref{tab:plda_comparison} shows that, on both the development and evaluation sets, the interpolated PLDA improves the diarization performance of AHC as compared to using PLDA trained only on out-of-domain VoxCeleb data. Still, the percentage of files where using the interpolated version worsens the results is 32\% for the development set and 45\% for evaluation.
However, the result on development data with the interpolated PLDA is overoptimistic as the test data are also used for the PLDA training. 

Note again that, in our diarization pipeline, the AHC is only used to obtain initial labels for the following BHMM based clustering. For more detailed analysis of the AHC-based diarization subsystem, we refer the reader to~\cite{DiezICASSP20}.


\begin{table}[htb]
\centering
\begin{tabular}{c|cc||cc}
  & VoxCeleb & Interpolated & Same & Improved\\ \hline
\multicolumn{1}{c|}{dev} & 20.46 & \textcolor{gray}{19.74} & \textcolor{gray}{9\%} & \textcolor{gray}{59\%}\\
\multicolumn{1}{c|}{eval} & 21.12 & 20.96 & 11\% & 45\%\\
\end{tabular}
\vspace*{-1.5mm}
\caption{DER on development and evaluation sets for AHC diarization using the PLDA trained on VoxCeleb and domain-adapted interpolated PLDA. Also percentage of files with equal or improved DER (when using the interpolated PLDA).}
\label{tab:plda_comparison}
\end{table}

\vspace*{-5.5mm}
\subsection{Bayesian HMM for x-vector clustering}
\label{sec:BHMM}
\vspace*{-1mm}
BHMM is used to cluster x-vectors as described in detail in~\cite{DiezInter19}.
Before the BHMM-based clustering, the x-vectors as well as the parameters of the PLDA model are projected using linear discriminant analysis (LDA). The LDA projection is calculated directly from the parameters of the original (interpolated) PLDA model described above. More details in \cite{landini2019but}. The resulting PLDA model is used in the BHMM to model speaker distributions as described in~\cite{DiezInter19}.

Variational Bayes (VB) inference for BHMM needs initial assignment of x-vectors to speaker clusters. This is taken from the previous AHC step, which needs to be run so as to under-cluster the x-vectors. This way the VB inference has more freedom to search for the optimal results and potentially remove redundant speakers\footnote{Note that the inference in BHMM cannot converge to higher number of speakers than what is suggested by the AHC-based initialization.}. A more thorough analysis on this matter is presented in \cite{DiezICASSP20}.

Iterative VB inference is run until convergence to update the assignment of x-vectors to speaker clusters. Automatic relevance determination (ARD)~\cite{Bishop2006} inherent in BHMM results in dropping redundant speaker clusters and allows us to properly estimate the number of speakers in each recording. 

To optimize the diarization performance, the VB inference is controlled by a number of tunable parameters.
In~\cite{DiezInter19}, we have newly introduced the \textit{speaker regularization coefficient $F_B$}, which affects the model to be more or less aggressive when dropping the redundant speakers. \textit{Acoustic scaling factor $F_A$} is introduced to compensate for the incorrect assumption of statistical independence between observations (i.e. x-vectors). $P_\text{loop}$ is the probability of not changing speakers  between observations, which serves as speaker turn duration model.
Note that our system uses a higher frame-rate for x-vector extraction compared to the ones used in former works, which requires using higher value of $P_\text{loop}$ and lower value of $F_A$ as compared to the optimal values reported in~\cite{DiezInter19}.
Details about the specific parameter values used for the challenge can be found in \cite{landini2019but}.

\begin{table}[htb]
\centering
\begin{tabular}{c|cc||cc}
  & VoxCeleb & Interpolated & Same & Improved\\ \hline
\multicolumn{1}{c|}{dev} & 18.34 & \textcolor{gray}{17.90} & \textcolor{gray}{14\%} & \textcolor{gray}{60\%} \\
\multicolumn{1}{c|}{eval} & 19.14 & 18.39 & 22\% & 56\% \\
\end{tabular}
\vspace*{-1.5mm}
\caption{DER on dev and eval sets for BHMM-based diarization using the PLDA trained on VoxCeleb and the domain-adapted interpolated PLDA. Also percentage of files with equal or improved DER (when using the interpolated PLDA).}
\label{tab:bhmm_comparison}
\end{table}

\vspace*{-1mm}
Table~\ref{tab:bhmm_comparison} shows the improvement obtained by using the BHMM model for diarization. 
When using the PLDA model trained on VoxCeleb data, the BHMM approach improves around 2\% absolute DER on both development and evaluation set with regard to the AHC (Table~\ref{tab:plda_comparison}). 
Moreover, by using the PLDA interpolation only 26\% of files in development and 22\% in evaluation have worse results than when using the PLDA trained on VoxCeleb, showing even better improvement than with AHC-based diarization.

\vspace*{-1mm}
\subsection{BHMM-based Re-segmentation}
\label{sec:resegmentation}
\vspace*{-1mm}
Given that the output from the previous step has a time resolution of 0.25s, we apply an additional frame-level VB re-segmentation step similar to the one used in our previous challenge submission~\cite{diez2018but}. 
VB re-segmentation is based on another BHMM which operates on mel-frequency cepstral coefficients (MFCC) with 10 ms frame rate as the input observations. In this case, state distributions are modeled by an i-vector extractor like model (i.e Gaussian mixture models with parameters constrained by eigenvoice priors)~\cite{DiezOdyssey18,KennyBayDiar}, which is pre-trained on VoxCeleb 2 data.
The initial assignments of MFCC frames to HMM states (speakers) is derived from the previous x-vector clustering step. Note that in the re-segmentation step only one VB iteration is performed~\cite{DiezInter19} instead of running the algorithm until convergence. 

\begin{table}[htb]
\centering
\begin{tabular}{c|cc}
  & BHMM & + reseg. \\ \hline
\multicolumn{1}{c|}{dev} & \textcolor{gray}{17.90} & \textcolor{gray}{18.23} \\
\multicolumn{1}{c|}{eval} & 18.39 & 18.38 \\
\end{tabular}
\vspace*{-1.5mm}
\caption{DER obtained using only BHMM at x-vector level and adding BHMM at frame-level (re-segmentation) on the development and evaluation sets.}
\label{tab:vb_comparison}
\end{table}

The effect of this extra re-segmentation step is shown in Table~\ref{tab:vb_comparison}. Unlike in our previous participation~\cite{diez2018but}, where the frame level re-segmentation gave great gains over simply doing AHC, this time this approach gives marginal gains. The reason for this is several-fold: the x-vectors are newly extracted every 0.25s, which provides 3 times better time resolution as compared to the typical 0.75s. Further, the BHMM-based x-vector clustering step produces a better diarization output than the AHC, leaving less margin for improvements. More importantly, BHMM-based x-vector clustering uses a PLDA model adapted to the target domain, whereas the re-segmentation step uses models solely trained on VoxCeleb data\footnote{This also explains the degradation obtained with re-segmentation step on dev data in Table~\ref{tab:vb_comparison}}. Adapting also the re-segmentation BHMM (i.e. the built-in i-vector model) would likely improve the re-segmentation results. See~\cite{DiezICASSP20} for more details.

\vspace*{-1mm}
\subsection{Overlapped Speech Post-processing}
\label{sec:overlap}
\vspace*{-1mm}
Given that none of our models accounts for overlapped speech (i.e. they all assume one speaker speaking at a time), we perform overlapped speech detection and apply a heuristic to label segments for more than one speaker.

For each recording, silence segments are removed and speech segments are concatenated. Then x-vectors are extracted from 1.5s sub-segments every 0.25s and classified using a logistic regression classifier as overlapped or non-overlapped speech. The classifier is trained on x-vectors extracted from the development set and labeled as overlapped speech if more than half of the original segment contains overlapped speech. Once overlap segments are detected, the heuristic consists in assigning for each frame in an overlapped speech segment the two closest speakers (in time) according to the diarization labels given by the previous step.


\begin{table}[htb]
\centering
\begin{tabular}{c|cc}
  & No ov. proc. & With ov. proc. \\ \hline
\multicolumn{1}{c|}{dev}  & \textcolor{gray}{18.23} & \textcolor{gray}{18.02} \\
\multicolumn{1}{c|}{eval} & 18.38 & 18.21
\end{tabular}
\vspace*{-1.5mm}
\caption{DER before and after doing overlapped speech post-processing on development and evaluation sets.}
\label{tab:overlap_comparison}
\end{table}

Table~\ref{tab:overlap_comparison} shows the comparison of results with and without overlap post-processing.
Although the x-vector extractor is trained with single speaker speech, the x-vectors still capture relevant information for overlapped speech detection. 
Note that 18.21 is slightly better than the result from our winning system\footnote{18.42 as shown in http://dihard.ldc.upenn.edu/competitions/73\#results}.
In our submission, a more complex PLDA adaptation scheme was used: Directions with larger within- and across-class variability in the in-domain PLDA model (trained on DIHARD dev) than in the out-of-domain PLDA (trained on VoxCeleb) were identified and the extra variability was added to the corresponding covariance matrices in the out-of-domain PLDA so that they would have at least as much variability as in the in-domain PLDA model. Later, we found that the simple interpolation of the two PLDA models is sufficient and even slightly improves the results.


\vspace*{-1mm}
\subsection{Voice Activity Detection}
\label{sec:vad_dihard}
\vspace*{-1mm}
Track 2 consisted in evaluating the same DIHARD set without using oracle VAD labels.
Although we explored using the same VAD system we had used before~\cite{diez2018but}, we found out that a DNN-based VAD system trained on the development set for binary, speech/non-speech, classification of speech frames provided better performance. More details in \cite{landini2019but}.


The model proposed for this track is essentially the same as for track 1 using the estimated (instead of oracle) VAD labels. However, due to a lack of time, PLDA interpolation is not done (the model trained on VoxCeleb data is used) and no overlapped speech post-processing is applied. 
With the DNN-based VAD, we obtain 23.81 DER on the development set (``cheating'') and 27.11 DER on the evaluation set.

\vspace*{-1mm}
\section{SYSTEMS for Tracks 3 and 4, CHIME DATA}
\label{sec:chime}
\vspace*{-2mm}
Tracks 3 and 4 proposed for the first time a multi-channel diarization task using CHiME 5 data. This posed new challenges with respect to performing diarization on the DIHARD data as we had to devise a system capable of using data collected in 4 channels arranged in microphone arrays. For these tracks we propose a much simpler diarization system.

\vspace*{-1mm}
\subsection{Multi-channel AHC Clustering}
\label{sec:track3}
\vspace*{-1mm}
Our diarization system is based on performing  AHC of x-vectors using information from all channels.
We analyzed two ways of taking advantage of multiple channels. One possibility we explored is to apply beamforming~\cite{anguera2006beamformit} using the four signals in order obtain a single beamformed channel to use as input for the diarization system. The second approach consists in extracting x-vectors from each channel, computing the corresponding pairwise similarity PLDA score matrices, averaging them and then performing AHC on the resulting score matrix.

For all approaches, the AHC of x-vectors is performed in a similar manner as for track 1: recordings from all channels are processed with the WPE method, x-vectors are extracted as described in \ref{sec:xvecs} and AHC is performed on the pairwise-similarity PLDA score matrices as in \ref{sec:initialization}. In this case, the PLDA model trained on VoxCeleb segments is adapted in an unsupervised way to the train and development data from the CHiME corpus. 

Table~\ref{tab:chime_channels} shows results for the different approaches: performing x-vector AHC on each channel separately, on the beamformed signal and on the PLDA score matrix obtained from the fusion of the PLDA matrices of each channel. Even though we found that the optimal thresholds for the different channels are different, all values correspond to the same threshold for the sake of comparison. As can be seen, there are some differences on performance when different channels are evaluated (up to 0.83 DER).

Using the beamformed signal does not provide better results than simply any of the single channels; still, the parameters used for producing the beamformed signal might not be optimal. Tuning of parameters was not explored during the challenge and remains as future work.

However, fusing the score matrices does improve the results. This pattern was seen for every threshold; however, the best AHC threshold for the \emph{Fusion} approach is always closer to 0, suggesting that the score fusion provides better calibrated scores. Note that the DER on the evaluation set for the \emph{Fusion} approach shown in the table is worse than our winning system\footnote{45.65 as shown in http://dihard.ldc.upenn.edu/competitions/75\#results}. 
This is because a different AHC threshold was used for the submission:
Since we saw different thresholds impacted considerably the performance in the dev+train set, we decided to submit diarization outputs obtained with different thresholds for evaluation expecting the performance would also vary.

\begin{table}[htb]
\setlength{\tabcolsep}{4pt} 
\centering
\begin{tabular}{c|cccccc}
& CH1 & CH2 & CH3 & CH4 & Beam. & Fusion \\ \hline
\multicolumn{1}{c|}{dev+train} & 55.43 & 55.34 & 55.78 & 54.95 & 55.75 & 53.58 \\
\multicolumn{1}{c|}{eval} & 48.55 & 48.37 & 48.19 & 48.3 & 50.31 & 47.93 \\
\end{tabular}
\vspace*{-1.5mm}
\caption{DER of our system on each channel separately, on the beamformed channel and when using fusion of scores, on the development+training and evaluation sets.}
\label{tab:chime_channels}
\end{table}

\vspace*{-5.5mm}
\subsection{Voice Activity Detection}
\vspace*{-1mm}
\label{sec:vad_chime}

For the track 4 of the challenge, the oracle VAD labels are not available and a VAD system~\cite{Matejka:LRE17} is run only on one of the channels to produce VAD labels.

With the NN-based VAD and the best aforementioned system, we obtain 58.92 DER which compares to 45.65 DER when oracle VAD labels are used. Training a tailored VAD system on the training set analogously as in track 2 remains as future work. 



\vspace*{-1mm}
\section{CONCLUSIONS}
\label{sec:conclusions}
\vspace*{-2mm}

Another edition of the DIHARD challenge has taken place and it has proven to be too hard for McClane\footnote{One day before the deadline, putting our faith in our latest system allowing us to obtain the first position in the ranking, we named the submission ``McClane''. However, a literally last minute submission allowed us to even improve its performance http://dihard.ldc.upenn.edu/competitions/73\#results}. One year has passed, the quality of x-vector extractors has improved and they have become the cornerstone for top-performing diarization systems. However, we have shown that it is possible to take more advantage of them when using BHMM in comparison to simply doing AHC. Given the current performance of the systems, the overlapped speech gains more relevance accounting for more than 50\% of the DER in our best systems in both sets. We believe this has to be addressed in the future as well as other approaches for adapting x-vectors to in-domain data. The challenge gave us the opportunity to work for the first time on multi-channel diarization. 

\label{sec:refs}

\bibliographystyle{IEEEbib}
\bibliography{biblio}

\begin{thebibliography}{10}

\bibitem{ryant2019second}
N.~Ryant, K.~Church, C.~Cieri, A.~Cristia, J.~Du, S.~Ganapathy, and
  M.~Liberman, ``The second dihard diarization challenge: Dataset, task, and
  baselines,'' {\em arXiv preprint arXiv:1906.07839}, 2019.

\bibitem{ryant2018first}
N.~Ryant, K.~Church, C.~Cieri, A.~Cristia, J.~Du, S.~Ganapathy, and
  M.~Liberman, ``First dihard challenge evaluation plan,'' 2018.

\bibitem{barker2018fifth}
J.~Barker, S.~Watanabe, E.~Vincent, and J.~Trmal, ``The fifth 'chime' speech
  separation and recognition challenge: Dataset, task and baselines,'' {\em
  arXiv preprint arXiv:1803.10609}, 2018.

\bibitem{landini2019but}
F.~Landini, S.~Wang, M.~Diez, L.~Burget, P.~Mat\v{e}jka,
  K.~\v{Z}mol\'{i}kov\'{a}, L.~Mo\v{s}ner, O.~Plchot, O.~Novotn\'{y},
  H.~Zeinali, and J.~Rohdin, ``{BUT System Description for DIHARD Speech
  Diarization Challenge 2019},'' {\em arXiv preprint arXiv:1910.08847}, 2019.

\bibitem{DIHARDrecipeBUT}
L.~Burget, M.~Diez, S.~Wang, and F.~Landini, ``{VBHMM x-vectors Diarization
  (aka VBx)}.''
  \url{https://speech.fit.vutbr.cz/software/vbhmm-x-vectors-diarization}.

\bibitem{DiezInter19}
M.~Diez, L.~Burget, S.~Wang, J.~Rohdin, and H.~\v{C}ernock\'{y}, ``{Bayesian
  HMM based x-vector clustering for Speaker Diarization},'' in {\em Proceedings
  of Interspeech}, 2019.

\bibitem{DiezIEEE19}
M.~Diez, L.~Burget, F.~Landini, and H.~\v{C}ernock\'{y}, ``Analysis of speaker
  diarization based on bayesian hmm with eigenvoice priors,'' {\em IEEE/ACM
  Transactions on Audio, Speech, and Language Processing}, 2019.

\bibitem{macartney2018improved}
C.~Macartney and T.~Weyde, ``Improved speech enhancement with the wave-u-net,''
  {\em arXiv preprint arXiv:1811.11307}, 2018.

\bibitem{plchot-enhancement-2016}
O.~Plchot, L.~Burget, H.~Aronowitz, and P.~Mat{\v{e}}jka, ``{A}udio {E}nhancing
  {W}ith {DNN} {A}utoencoder {F}or {S}peaker {R}ecognition,'' in {\em
  Proceedings of the 41th IEEE International Conference on Acoustics, Speech
  and Signal Processing (ICASSP 2016), 2016}, pp.~5090--5094, IEEE Signal
  Processing Society, 2016.

\bibitem{nakatani-wpe-2010}
T.~Nakatani, T.~Yoshioka, K.~Kinoshita, M.~Miyoshi, and B.~H. Juang, ``Speech
  dereverberation based on variance-normalized delayed linear prediction,''
  {\em IEEE Transactions on Audio, Speech, and Language Processing}, vol.~18,
  pp.~1717--1731, Sept 2010.

\bibitem{Drude2018NaraWPE}
L.~Drude, J.~Heymann, C.~Boeddeker, and R.~Haeb-Umbach, ``{NARA-WPE: A Python
  package for weighted prediction error dereverberation in Numpy and Tensorflow
  for online and offline processing},'' in {\em 13. ITG Fachtagung
  Sprachkommunikation}, Oct 2018.

\bibitem{diez2018but}
M.~Diez, F.~Landini, L.~Burget, J.~Rohdin, A.~Silnova,
  K.~\v{Z}mol{\'\i}kov{\'a}, O.~Novotn{\'y}, K.~Vesel{\'y}, O.~Glembek,
  O.~Plchot, {\em et~al.}, ``{BUT System for DIHARD Speech Diarization
  Challenge 2018},'' in {\em Proceedings of Interspeech 2018}, pp.~2798--2802,
  2018.

\bibitem{SellSDIHARD18}
G.~Sell, D.~Snyder, A.~McCree, D.~Garcia{-}Romero, J.~Villalba, M.~Maciejewski,
  V.~Manohar, N.~Dehak, D.~Povey, S.~Watanabe, and S.~Khudanpur, ``Diarization
  is hard: Some experiences and lessons learned for the {JHU} team in the
  inaugural {DIHARD} challenge,'' in {\em Interspeech}, pp.~2808--2812, {ISCA},
  2018.

\bibitem{DiezICASSP20}
M.~Diez, L.~Burget, F.~Landini, S.~Wang, and H.~\v{C}ernock\'{y}, ``{Optimizing
  Bayesian HMM based x-vector clustering for the second DIHARD speech
  diarization challenge},'' in {\em Proceedings of International Conference on
  Acoustics, Speech and Signal Processing (ICASSP)}, 2020.

\bibitem{snyder_kaldi_recipe}
Kaldi, ``{SRE16 v2}.''
  https://github.com/kaldi-asr/kaldi/tree/master/egs/sre16/v2.

\bibitem{povey2011kaldi}
D.~Povey, A.~Ghoshal, G.~Boulianne, L.~Burget, O.~Glembek, N.~Goel,
  M.~Hannemann, P.~Motli\v{c}ek, Y.~Qian, P.~Schwarz, {\em et~al.}, ``The kaldi
  speech recognition toolkit,'' in {\em IEEE 2011 workshop on automatic speech
  recognition and understanding}, IEEE Signal Processing Society, 2011.

\bibitem{chung2018voxceleb2}
J.~S. Chung, A.~Nagrani, and A.~Zisserman, ``Voxceleb2: Deep speaker
  recognition,'' {\em arXiv preprint arXiv:1806.05622}, 2018.

\bibitem{kenny10PLDA_HTP}
P.~Kenny, ``{Bayesian Speaker Verification with Heavy-Tailed Priors},'' in {\em
  in Proceedings of Odyssey}, June 2010.

\bibitem{GarciaRomero2011lnorm}
D.~Garcia-Romero and C.~Y. Espy-Wilson, ``Analysis of i-vector length
  normalization in speaker recognition systems,'' in {\em Proceedings of
  Interspeech 2011}, 2011.

\bibitem{Bishop2006}
C.~M. Bishop, {\em Pattern Recognition and Machine Learning (Information
  Science and Statistics)}.
\newblock Secaucus, NJ, USA: Springer-Verlag New York, Inc., 2006.

\bibitem{DiezOdyssey18}
M.~Diez, L.~Burget, and P.~Mat\v{e}jka, ``Speaker diarization based on bayesian
  hmm with eigenvoice priors,'' in {\em Proceedings of Odyssey 2018, The
  speaker and Language Recognition Workshop}, 2018.

\bibitem{KennyBayDiar}
P.~Kenny, ``Bayesian analysis of speaker diarization with eigenvoice priors,''
  tech. rep., Montreal: CRIM, 2008.

\bibitem{anguera2006beamformit}
X.~Anguera, ``Beamformit, the fast and robust acoustic beamformer,'' 2006.

\bibitem{Matejka:LRE17}
P.~Mat\v{e}jka and et.al., ``{BUT}-{PT} system description for nist lre,'' in
  {\em Proceedings of NIST Language Recognition Workshop 2017}, pp.~1--6,
  National Institute of Standards and Technology, 2017.

\end{thebibliography}

\end{document}